\documentclass[twocolumn,english,prb]{revtex4-1}
\usepackage[T1]{fontenc}
\usepackage[latin9]{inputenc}
\usepackage{amsmath}
\usepackage{graphicx}
\usepackage{amssymb}

\makeatletter
 \@ifundefined{textcolor}{}
 {%
   \definecolor{BLACK}{gray}{0}
   \definecolor{WHITE}{gray}{1}
   \definecolor{RED}{rgb}{1,0,0}
   \definecolor{GREEN}{rgb}{0,1,0}
   \definecolor{BLUE}{rgb}{0,0,1}
   \definecolor{CYAN}{cmyk}{1,0,0,0}
   \definecolor{MAGENTA}{cmyk}{0,1,0,0}
   \definecolor{YELLOW}{cmyk}{0,0,1,0}
 }

\makeatother

\usepackage{babel}

\makeatother

\usepackage{babel}

\begin{document}

\title{Normal state bottleneck and nematic fluctuations from femtosecond
quasi-particle relaxation dynamics in Sm(Fe,Co)AsO}

\author{T. Mertelj$^{1}$, L. Stojchevska$^{1}$, N. D. Zhigadlo$^{2}$,
J. Karpinski$^{2,3}$ and D. Mihailovic$^{1}$}

\affiliation{$^{1}$Complex Matter Dept., Jozef Stefan Institute, Jamova 39, Ljubljana,
SI-1000, Ljubljana, Slovenia }

\affiliation{$^{2}$Laboratory for Solid State Physics, ETH Zürich, 8093 Zürich,
Switzerland}

\affiliation{$^{3}$Institute of Condensed Matter Physics, EPFL, CH-1015 Lausanne,
Switzerland}

\date{\today}
\begin{abstract}
We investigate temperature and fluence dependent dynamics of the photoexcited
quasi-particle relaxation and low-energy electronic structure in electron-doped
1111-structure Sm(Fe$_{0.93}$Co$_{0.07}$)AsO single crystal. We
find that the behavior is qualitatively identical to the 122-structure
Ba(Fe,Co)$_{2}$As$_{2}$ including the presence of a normal state
pseudogap and a marked 2-fold symmetry breaking in the tetragonal
phase that we relate to the electronic nematicity. The 2-fold symmetry
breaking appears to be a general feature of the electron doped iron
pnictides.
\end{abstract}
\maketitle

\section{Introduction}

Recently, the presence of the normal-state electronic nematic fluctuations\cite{ChuAnalytis2010,ChuangAllan2010,TanatarBlomberg2010,DuszaLucarelli2011,YingWang2011,YiLu2011,StojchevskaMertelj2012,ZhangHe2012,MartinelliPalenzona2011,MartinelliPalenzona2012}
in layered iron-based pnictide superconductors\cite{kamiharaWatanabe2008}
has been suggested from a remarkable 2-fold anisotropy of physical
properties in the strained tetragonal phase, well above the structural
phase transition. The microscopic origin of the observed anisotropy
and possible relation to spin-density-wave (SDW) ordering and superconductivity
is still under intense debate. 

Most of the published experimental work on 2-fold normal state planar
anisotropy is based on 122-structure systems such as Co-doped BaFe$_{2}$As$_{2}$
(Ba-122) with only a few exceptions.\cite{ZhangHe2012,MartinelliPalenzona2011,MartinelliPalenzona2012}
Moreover, hole doped BaFe$_{2}$As$_{2}$\cite{YingWang2011} shows
only a tiny anisotropy in resistivity indicating that the domain of
the presence of the nematicity might be limited to the electron doped
compounds. It is therefore important to establish experimentally to
what extent the tendency for a 2-fold symmetry breaking is general
also \emph{for different members} of the electron doped iron-based
pnictide superconductors family.

Photoexcited carrier dynamics in iron-based pnictides\cite{MerteljKabanov2009prl,MerteljKabanov2009jsnm,MerteljKusar2010,TorchinskyChen2010,ChiaTalbayev2010,StojchevskaKusar2010,GongLai2010,MansartBoschetto2010,TorchinskyMcIver2011,RettigCortes2012}
was proven to be a useful probe of the electronic structure. Recently,
we demonstrated a remarkable sensitivity of the photoinduced reflectivity
transients to the 2-fold symmetry breaking in Co doped Ba-122.\cite{StojchevskaMertelj2012}
Here we extend our study to Co doped 1111-structure SmFeAsO (Sm-1111)
and find that behavior is qualitatively identical to the behavior
in Ba-122, with the clear presence of the 2-fold anisotropy up to
$\sim$170 K.

\section{Experimental}

\subsection{Setup and sample}

The SmFe$_{0.93}$Co$_{0.07}$AsO single crystal with $T{}_{\mathrm{c}}=15.2$
K ($T{}_{\mathrm{c,midpoint}}=14$ K) and several hundred micrometers
in size was grown from NaAs flux at ETH Zurich using high-pressure
and high-temperature technique.\cite{ZhigadloWeyeneth2012} For optical
measurements the crystal was glued onto a sapphire substrate and cleaved
by a razor blade before mounting in an optical liquid-He flow cryostat. 

Measurements of the photoinduced reflectivity, $\Delta R/R$, were
performed using the standard pump-probe technique, with 50 fs optical
pulses from a 250-kHz Ti:Al$_{2}$O$_{3}$ regenerative amplifier
seeded with an Ti:Al$_{2}$O$_{3}$ oscillator. We used the pump photons
with the doubled ($\hbar\omega_{\mathrm{P}}=3.1$ eV) photon energy
and the probe photons with the laser fundamental 1.55 eV photon energy
to easily suppress the scattered pump photons by long-pass filtering.
The pump and probe beams were nearly perpendicular to the cleaved
sample surface with polarizations perpendicular to each other and
oriented with respect to the crystals to obtain the maximum/minimum
amplitude of $\Delta R/R$ at low temperatures. The pump and probe
beam diameters at the sample position were 110 $\mu$m and 52$\mu$m,
respectively.

\begin{figure}[tbh]
\begin{centering}
\includegraphics[width=0.45\textwidth]{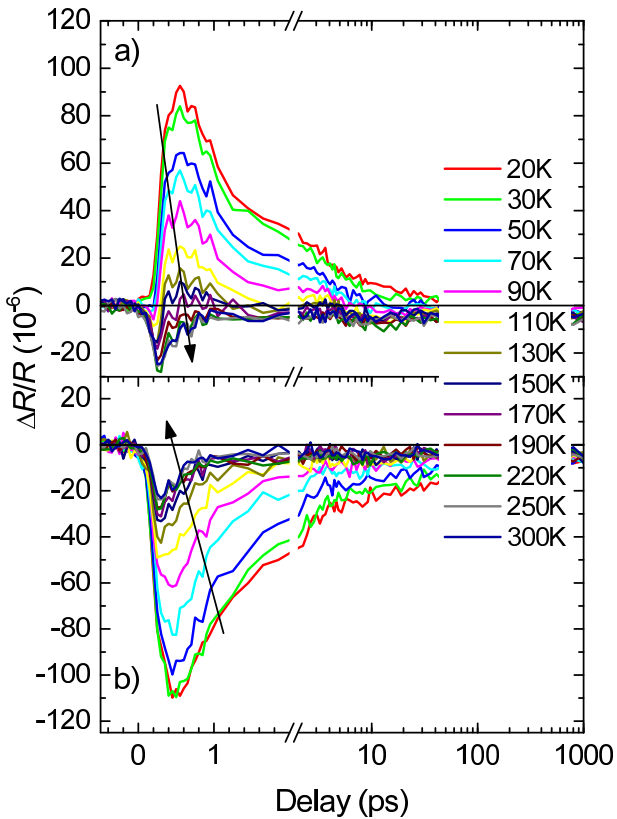} 
\par\end{centering}

\caption{(Color online) The raw $\Delta R/R$ transients as a function of temperature
at 8 $\mu$J/cm$^{2}$ pump fluence. a) and b) correspond to $\mathcal{P}^{+}$
and $\mathcal{P}^{-}$ polarizations, respectively. The arrows indicate
increasing temperature.}

\label{fig:fig-DRvsT} 
\end{figure}
\begin{figure}[tbh]
\begin{centering}
\includegraphics[width=0.45\textwidth]{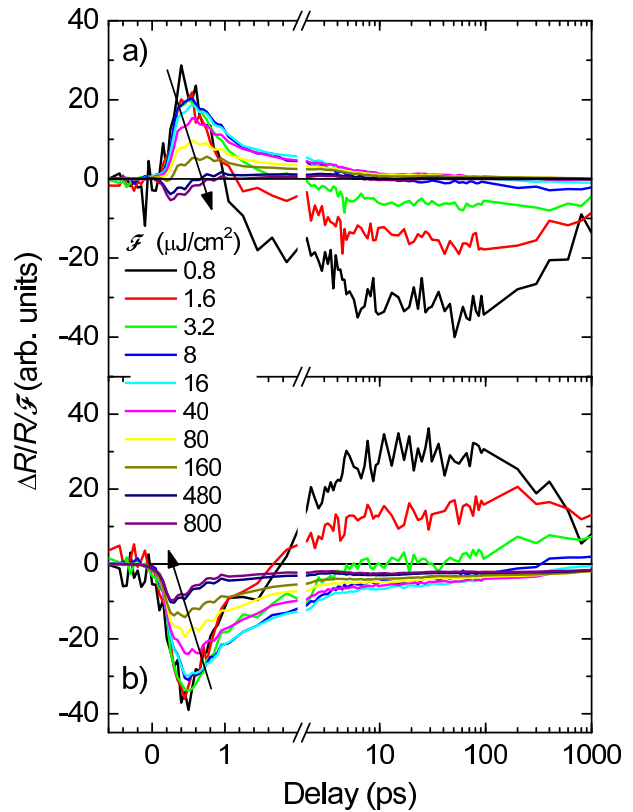} 
\par\end{centering}

\caption{(Color online) Fluence-normalized $\Delta R/R/\mathcal{F}$ transients
as a function of $\mathcal{F}$ at $T=5$ K. The arrows indicate the
direction of increasing $\mathcal{F}$. Overlap of the curves indicates
a linear $\mathcal{F}$ dependence. a) and b) correspond to the $\mathcal{P}^{+}$
and $\mathcal{P}^{-}$ polarization, respectively.}

\label{fig:fig-DR-vs-F} 
\end{figure}

\subsection{Overview of the experimental data set}

In Fig. \ref{fig:fig-DRvsT} we plot the temperature dependence of
the raw photoinduced reflectivity ($\Delta R/R$). In spite of the
fact, that no deliberate uniaxial strain was applied to the sample,
the transients develop a 2-fold rotational anisotropy with respect
to the probe polarization upon cooling. At the room temperature the
transients are nearly identical for both polarizations. Below $\sim200$
K the transients for different probe polarizations start to show different
time dependencies. The transients for one of the polarizations eventually
change the sign with decreasing temperature. In the absence of information
which crystallographic directions correspond to the two different
orthogonal probe polarizations we denote the polarization corresponding
to the low-temperature minimal and maximal sub-picosecond peak $\Delta R/R$
value $\mathcal{P}^{-}$ and $\mathcal{P}^{+}$, respectively.

At low temperatures the $\Delta R/R$ transients show a strong pump
fluence ($\mathcal{F}$) dependence. In the superconducting state,
at the lowest experimentally feasible fluence, the transients in both
probe polarizations change sign after the initial sub-ps peak followed
by a slow decay on several-hundred-ps timescale (see Fig. \ref{fig:fig-DR-vs-F}).
The amplitude of the slow part of the transients shows saturation
with increasing $\mathcal{F}$ already at the lowest $\mathcal{F}=0.8$
$\mu$J/cm$^{2}$, while the sub-ps peak magnitude is linear with
$\mathcal{F}$ up to $\mathcal{F}\simeq10$ $\mu$J/cm$^{2}$. With
further increase of $\mathcal{F}$ the amplitude of the sub-ps peak
also saturates and the signal evolves towards the shape observed at
the room temperature. At the room temperature the transients show
a linear increase of the magnitude with increasing $\mathcal{F}$
with no signs of any saturation or change in the relaxation time.

\begin{figure}[tbh]
\begin{centering}
\includegraphics[width=0.45\textwidth]{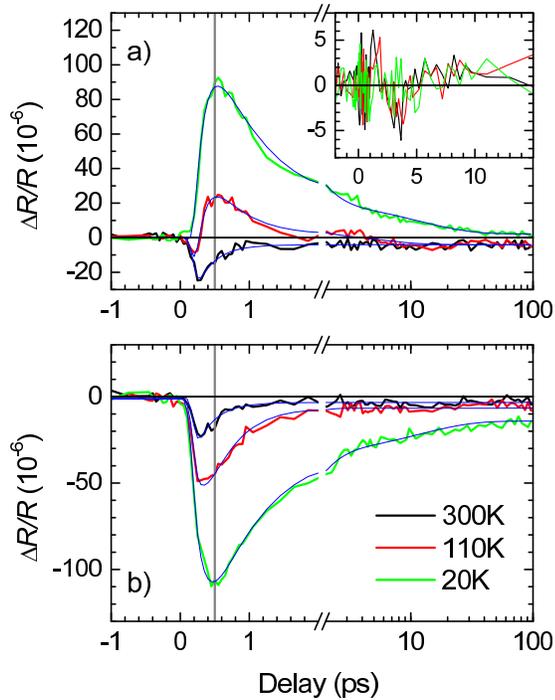} 
\par\end{centering}

\caption{(Color online) Examples of multi-exponent fits (\ref{eq:fitfunc})
in the normal state for both polarizations. The inset shows the systematic
fit error observed for the $\mathcal{P}^{+}$polarization. The vertical
lines represent the peaks of the low-$T$ transients at 0.5 ps.}

\label{fig:fig-fits} 
\end{figure}

\section{Discussion}

\subsection{Relaxation components in the normal state}

In order to determine $T$-dependencies of relaxation times we fit
the transients for both probe polarizations up to 100 ps delay with
three exponentially decaying components:\cite{MihailovicDemsar99}

\begin{eqnarray}
\frac{\Delta R}{R} & = & \underset{i\in\{\mathrm{A,B,C}\}}{\sum}\frac{A_{i}}{2}\mathrm{e}^{-\frac{t-t_{0}}{\tau_{i}}}\operatorname{erfc}\left(\frac{\sigma^{2}-4(t-t_{0})\tau_{i}}{2\sqrt{2}\sigma\tau_{i}}\right)+\nonumber \\
 &  & +\frac{A_{\mathrm{D}}}{2}\operatorname{erfc}\left(-\frac{\sqrt{2}(t-t_{0})}{\sigma}\right),\label{eq:fitfunc}\end{eqnarray}
where $\sigma$ corresponds to the effective width of the excitation
pulse with a Gaussian temporal profile arriving at $t_{0}$ and $\tau_{i}$
the exponential relaxation times. Component D represents the relaxation
on a timescale beyond 100 ps. During the fitting the relaxation times
for the two orthogonal polarizations were linked while amplitudes
were kept independent. Examples of fits are shown in Fig. \ref{fig:fig-fits}
for the three characteristic temperature regions. 

There is a systematic discrepancy between the fits and the data for
$\mathcal{P}^{+}$ indicating the presence of another weak component.
The fit error shown in the inset to Fig. \ref{fig:fig-fits} suggest
that this component is rather weak and temperature independent so
it will not be discussed further here.

\begin{figure}[tbh]
\begin{centering}
\includegraphics[width=0.45\textwidth]{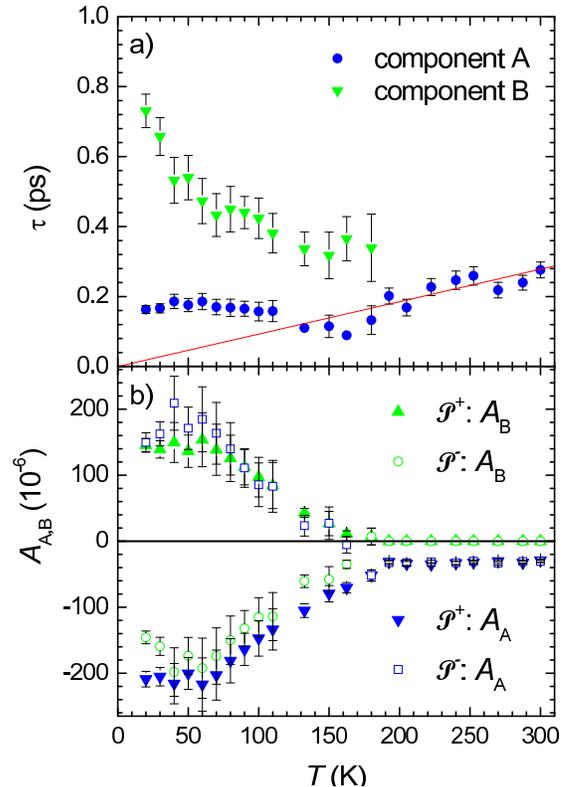} 
\par\end{centering}

\caption{(Color online) The fastest two multi-exponential fit relaxation times
a) and corresponding amplitudes b). The red (gray) line is a fit of
Eq. (\ref{eq:TauPoorHigh}) discussed in the text.}

\label{fig:fig-fA-vs-T} 
\end{figure}

\begin{figure}[tbh]
\begin{centering}
\includegraphics[width=0.4\textwidth]{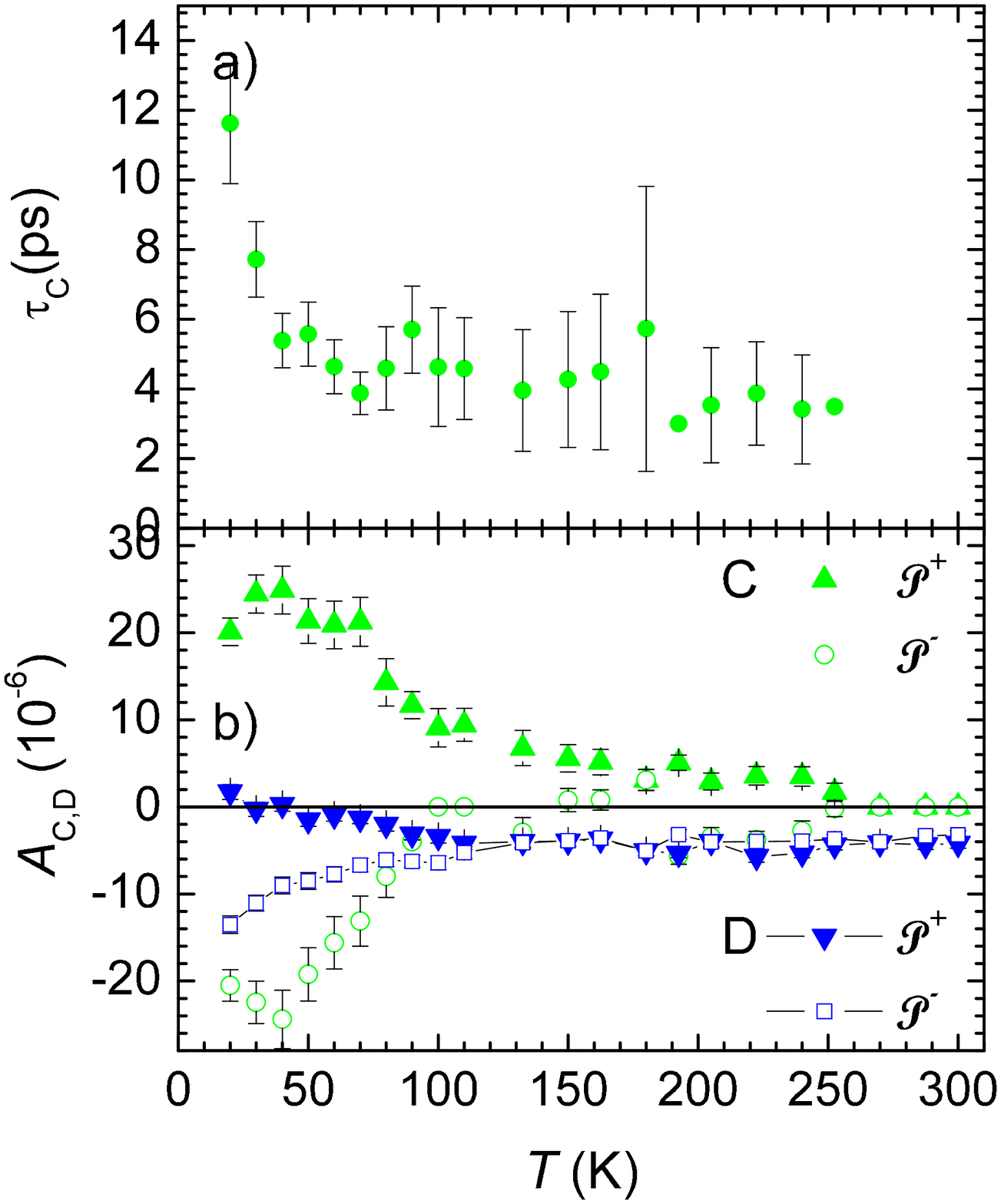} 
\par\end{centering}

\caption{(Color online) The slowest multi-exponential fit relaxation time a)
and the corresponding amplitude b) together with the long-delay residual
$A_{\mathrm{D}}$.}

\label{fig:fig-A-vs-T-slow} 
\end{figure}

At the highest temperatures a single exponential component named A,
with a linearly $T$-dependent relaxation time, $\tau_{\mathrm{A}}\sim0.2-0.3$
ps, completely describes the initial sub-ps relaxation (see Fig. \ref{fig:fig-fits}
and \ref{fig:fig-fA-vs-T}).

Below $\sim170$ K, where the anisotropy becomes clearly visible and
the $\mathcal{P}^{+}$- polarization component A changes sign, another
component named B appears with a slightly longer relaxation time and
a strong amplitude anisotropy. The relaxation time, $\tau_{\mathrm{B}}$,
{[}see Fig. \ref{fig:fig-fA-vs-T} (a){]} shows an increase with decreasing
$T$ from $\sim0.4$ ps at $\sim170$K to $\sim0.7$ ps at 5 K, while
$\tau_{\mathrm{A}}$ remains constant slightly below 0.2 ps in this
$T$-range. 

Component C {[}see Fig. \ref{fig:fig-A-vs-T-slow}{]} has the longest
relaxation time ($\tau_{\mathrm{C}}$). The component is absent at
the room $T$ and appears below $\sim250$ K. The amplitude is rather
small an $T$-independent down to $\sim120$ K and then increases
with decreasing temperature and saturates below $\sim70$ K. $\tau_{\mathrm{C}}$
is, differently from the amplitude, temperature independent above
$\sim70$ K while it steeply increases with decreasing $T$ in the
region where the amplitude saturates. 

The amplitude of component D is also anisotropic at the lowest $T$.
With increasing $T$ the anisotropy decreases and vanishes above $120$
K, were the amplitude becomes $T$-independent. 

\begin{figure}[tbh]
\begin{centering}
\includegraphics[bb=0bp 200bp 595bp 750bp,clip,width=0.45\textwidth]{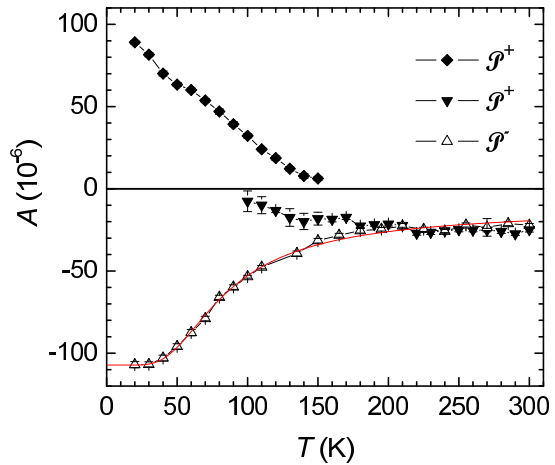} 
\par\end{centering}

\caption{(Color online) The magnitudes of $\Delta R/R$ transients extrema
as a function of $T$. The red (gray) lines is the fit of Eq. (\ref{eq:AvsT-PG})
discussed in the text.}

\label{fig:fig-A-vs-T} 
\end{figure}

\subsection{Pseudogap and 2-fold in-plane anisotropy}

Above $T_{\mathrm{c}}$ the amplitudes of the transients show a gradual
decrease with increasing temperature (see Fig. \ref{fig:fig-A-vs-T}).
Such a decrease can be associated with a bottleneck in relaxation
and therefore the presence of a pseudogap in the quasiparticle density
of states. To obtain a quantitative information about the pseudogap
we analyze the normal state $T$-dependent amplitudes in the context
of the relaxation across a $T$-independent gap.\cite{KabanovDemsar99,MerteljKabanov2009prl,MerteljKusar2010}:
\begin{alignat}{1}
\mbox{\ensuremath{\Delta}}R\propto & n_{\mathrm{pe}}\propto\left[1+g_{\mathrm{ph}}\exp\left(-\frac{\Delta_{\mathrm{PG}}\left(T\right)}{k_{\mathrm{B}}T}\right)\right]^{-1},\label{eq:AvsT-PG}\end{alignat}
where $g_{\mathrm{ph}}$ is the ratio between the number of involved
phonons and number of involved quasi-particle states.\cite{KabanovDemsar99}
As shown in Fig. \ref{fig:fig-A-vs-T} the $T$-dependent amplitude
for the $\mathcal{P}^{-}$ polarization can be well fit using equation
(\ref{eq:AvsT-PG}) with $2\Delta_{\mathrm{PG}}=440$ K. This value
is slightly lower than in Co doped Ba-122.\cite{StojchevskaMertelj2012}

On the other hand, the $T$-dependent transients for the $\mathcal{P}^{+}$
polarization show the change of sign with increasing $T$ including
a non-monotonous relaxation in the 90-170 K range and can not be described
directly by equation (\ref{eq:AvsT-PG}).%
\footnote{The behavior is similar to the Co doped Ba-122,\cite{StojchevskaMertelj2012}
but the roles of polarizations are reversed presumably due to slightly
different optical matrix elements.%
} The failure of Eq. (\ref{eq:AvsT-PG}) for description of the $\mathcal{P}^{+}$
polarization amplitude $T$-dependence, however, does not contradict
to the bottleneck. Eq. (\ref{eq:AvsT-PG}) does not take into account
the intrinsic multi-band nature of iron-pnictides leading to the multi
component relaxation and a possible $T$-dependence of the optical-transition
matrix elements. Indeed, the $T$-dependent amplitudes of each component
separately (Fig. \ref{fig:fig-fA-vs-T} (b)) are consistent with the
bottleneck scenario while the conspicuous sign change of $\mathcal{P}^{+}$-polarization
component A amplitude at $\sim170$ K can be attributed to the change
of the optical matrix elements due to the band shifts as discussed
already for the SDW state in Co doped Ba-122.\cite{StojchevskaMertelj2012}

Comparing the relaxation times of the strongest two relaxation components
with time resolved ARPES\cite{RettigCortes2012} in Eu-122 shows a
striking similarity in both magnitudes and temperature dependencies.
According to ARPES different relaxation times correspond to different
regions in the Brillouine zone, where in the SDW state the slower
relaxation, with $\tau\sim0.8$ ps, is due to the relaxation of the
photexcited electrons around the $\Gamma$ point and the fastest,
with $\tau\sim0.15$ ps corresponds to the photexcited holes away
from the the $\Gamma$ point. Since our sample shows no SDW ordering
the similarity suggests that the pseudogap has a similar origin as
the charge gap observed in the SDW state.

As in the case of Co-doped Ba-122\cite{StojchevskaMertelj2012} the
most striking feature of our data set is the observation of the 2-fold
in-plane rotational anisotropy of the optical transients despite the
tetragonal lattice symmetry (see Fig. \ref{fig:fig-DRvsT}). In the
absence of any structural data\cite{KhanRahman2012} which would indicate
that our sample is not tetragonal in the thermodynamic equilibrium
we assume that the breaking of the 4-fold tetragonal symmetry is not
spontaneous, but is a consequence of an external anisotropy such as
anisotropic boundary and/or excitation conditions that introduce an
anisotropic surface strain.\cite{StojchevskaMertelj2012} Since in
our experiment any external anisotropy must be weak our observations
suggest that the intrinsic susceptibility for 2-fold symmetry breaking
is large and increases with decreasing $T.$

Using the same arguments as in the case of Co doped Ba-122\cite{StojchevskaMertelj2012}
we tentatively associate the 2-fold symmetry breaking instability
with electronic nematic fluctuations\cite{ChuAnalytis2010} or ordering
of the Fe $d$ orbitals. Due to the concurrent appearance of the bottleneck
and the increase of the polarization anisotropy of the transients
during cooling the pseudogap might be associated with the 4-fold rotational
symmetry breaking. 

Our observations are consistent with the recently observed\cite{MartinelliPalenzona2011,MartinelliPalenzona2012}
weak structural tetragonal symmetry breaking above $T\simeq140$ K
in F doped Sm-1111 by a high resolution synchrotron powder diffraction,
extending well into the SC dome region of the phase diagram.

\subsection{Electron phonon coupling }

Since our data suggest that the effects of the nematic fluctuations
become negligible above $T\simeq200K$ we analyze the initial part
of the $\Delta R/R$ transients at high $T$ in the framework of the
electron-phonon relaxation in metals.\cite{KabanovAlexandrov2008,GadermaierAlexandrov2010}
The $\mathcal{F}$-independent relaxation at the room temperature
warrants use of the low excitation expansion,\cite{KabanovAlexandrov2008}
where in the high temperature limit the energy relaxation time is
proportional to $T$,\cite{KabanovAlexandrov2008,GadermaierAlexandrov2010}\begin{equation}
\tau=\frac{2\pi k_{\mathrm{B}}T}{3\hbar\lambda\langle\omega^{2}\rangle}.\label{eq:TauPoorHigh}\end{equation}
Here $\lambda\langle\omega^{2}\rangle$ is the second moment of the
Eliashberg function, $\alpha^{2}F(\omega)$, and $k_{\mathrm{B}}$
the Boltzman constant.\cite{KabanovAlexandrov2008}.

The temperature dependence of component A relaxation time shows a
clear linear $T$-dependence predicted by equation (\ref{eq:TauPoorHigh})
above $T\simeq200K$ corresponding to $\lambda\langle\omega^{2}\rangle=130$
meV$^{2}$, which is the same as in the undoped Sm-1111\cite{MerteljKusar2010}
and suggests a moderate electron phonon coupling constant $\lambda\sim0.2.$

\subsection{Superconducting state}

\begin{figure}[tbh]
\begin{centering}
\includegraphics[bb=60bp 0bp 530bp 842bp,clip,width=0.45\textwidth]{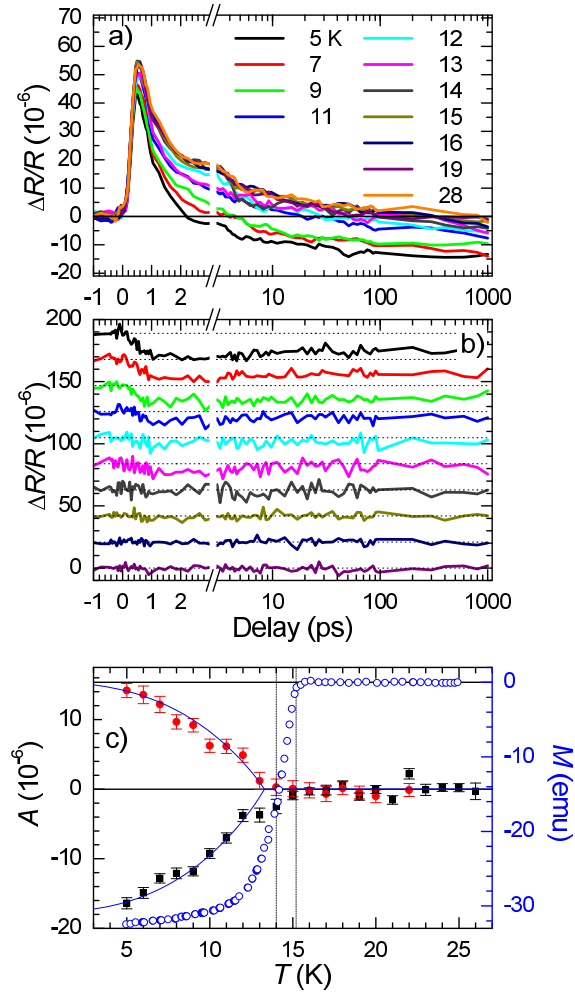} 
\par\end{centering}

\caption{(Color online) The raw $\Delta R/R$ transients as a function of $T$
at 3.2 $\mu$J/cm$^{2}$ pump fluence for the $\mathcal{P}^{+}$ polarization
(a). The superconducting response for the $\mathcal{P}^{+}$ polarization
obtained by subtraction of the response above $T_{\mathrm{c}}$ (b).
The traces are shifted vertically for clarity as indicated by the
dotted lines. The magnitude of the SC response for $\mathcal{P}^{+}$
and $\mathcal{P}^{-}$ polarizations as functions of $T$ compared
to the SQUID magnetometry data (c). The lines in (c) are Mattis-Bardeen
fits (\ref{eq:MattBard}) discussed in text. The vertical dashed lines
indicate $T_{\mathrm{c}}$ and $T\mathrm{_{c,midpoint}}$ obtained
from the SQUID susceptibility data.}

\label{fig:fig-DR-vs-T-SC} 
\end{figure}

The slowest part of the relaxation shows a strong temperature and
fluence dependencies below $T_{\mathrm{c}}$. Since at low $\mathcal{F}$
there is virtually no $T$-dependence of the transients above $T_{\mathrm{c}}=15.2$
K up to $\sim30$K (see Fig. \ref{fig:fig-DR-vs-T-SC}) we extract
the slow superconducting (SC) component by subtraction of the averaged
transients measured just above $T_{\mathrm{c}}$, as shown in Fig.
\ref{fig:fig-DR-vs-T-SC} (b). At $\mathcal{F}=3.2$ $\mu$J/cm$^{2}$
the SC component has a risetime of $\sim$1 ps and shows significantly
slower decay time than our time window. Consistently with previous
observations in the cuprates and iron-based pnictides\cite{KusarKabanov2008,StojchevskaKusar2011,CoslovichGiannetti2011}
the SC component is associated with a complete photodestruction of
the SC state. 

We fit the temperature dependence of the saturated SC-component amplitude
(see inset to Fig. \ref{fig:fig-DR-vs-T-SC}) using the high-frequency
limit of the Mattis-Bardeen formula,\cite{MattisBardeen1985,MerteljKusar2010}
\begin{equation}
\frac{\Delta R}{R}\propto\left(\frac{\Delta\left(T\right)}{\hbar\omega}\right)^{2}\log\left(\frac{3.3\hbar\omega}{\Delta\left(T\right)}\right),\label{eq:MattBard}\end{equation}
where $\hbar\omega$ is the probe-photon energy and $\Delta\left(T\right)$
the superconducting gap with the BCS temperature dependence. From
the fit we obtain $T_{\mathrm{c}}=13.3$ K which is 1.9 K smaller
than $T_{\mathrm{c}}$ from the SQUID magnetometry data. Due to the
small signal to noise level of the SC component it is not possible
to reliably determine whether this discrepancy originates from the
laser heating of the excited volume%
\footnote{Due to the laser heating the temperature of the excited spot on the
sample is higher than the cryostat temperature leading to an apparent
decrease of $T_{\mathrm{c}}$.%
} or the sample is inhomogeneous with variations of $T_{\mathrm{c}}$
across the sample as suggested by the width of the susceptibility
transition with $T_{\mathrm{c}}-T_{\mathrm{c,midpint}}=1.2$ K.

Moreover, the SC component shows a similar anisotropy as the normal
state components. The presence of the anisotropy concurrent with the
presence of transients due to the pseudogap bottleneck below $T_{\mathrm{c}}$
indicate that the superconductivity is coexisting with nematicity
and the pseudogap.

\section{Summary and conclusions}

We performed a time-resolved optical spectroscopy study in near optimally
doped Sm(Fe$_{0.93}$Co$_{0.07}$)AsO pnictide superconductor. As
in all previously studied electron doped iron pnictide superconductors\cite{MerteljKusar2010,StojchevskaMertelj2012}
we find a normal state bottleneck in the relaxation associated with
opening of a pseudogap in the electronic density of states and a moderate
electron phonon coupling constant.

Due to the observation of a marked 2-fold in-plane anisotropy below
$\sim170$ K, that is concurrent with the appearance of the pseudogap,
we tentatively associate the anomalous normal state properties with
electronic nematic fluctuations, similar as found in Ba(Fe,Co)$_{2}$As$_{2}$.
The high susceptibility towards 2-fold symmetry breaking therefore
seems to be a general feature of electron doped (Fe,As) planes. The
presence of the anisotropy in the superconducting state indicates
a coexistence of superconductivity and the nematic electronic order
in these compounds.
\begin{acknowledgments}
Work at Jozef Stefan Institute was supported by ARRS (Grant No. P1-0040).
The work at ETH Zürich was supported by Swiss National Science Foundation
and the National Center of Competence in Research MaNEP. J. K. acknowledges
support from ERC project Super Iron.
\end{acknowledgments}
\bibliography{biblio}

\end{document}